# Exploring dark photons via a subfrequency laser search in gravitational wave detectors

M. Afif Ismail[,1,2] Chrisna Setyo Nugroho[,3,*] and Henry Tsz-King Wong[1]

[1]*Institute of Physics, Academia Sinica, Taipei 11529, Taiwan*
[2]*Department of Physics, National Central University, Chungli 32001, Taiwan*
[3]*Department of Physics, National Taiwan Normal University, Taipei 116, Taiwan*

We propose a novel idea to detect a dark photon in gravitational wave experiments. Our setups are capable of performing the whole process of dark photon production, its decay products, and new physics signal discovery. This "mini-LHC" is inspired by the recent idea of dark photon detection using laser light in light shining through the wall (LSW) experiments such as ALPS II. Taking the subfrequency light emitted from the laser source as the new physics signal, we show that the sensitivity of our proposal is 2 orders of magnitude better than the original idea in the LSW studies.

## I. INTRODUCTION

The existence of nonluminous matter dubbed dark matter (DM) provides explanations for several astronomy observations, e.g., rotational curves of galaxies. However, except for its gravitational effects, the detailed properties, including its spin and mass, remain unknown. The DM particle could be a boson or a fermion, and its mass could be as light as below eV or it could be heavier than TeV. Even though no conclusive evidence of a DM signal has been found in the laboratory, some observations, if interpreted as signals of DM, give us hints about the mass of DM [1–10].

As an extension of the Standard Model of particle physics (SM), dark photon $\gamma'$ and axionlike particle (ALP) $a$ provide suitable DM candidates (see the recent review [11,12] and the references therein). Both the dark photon and ALP interact with the photon. The former communicates with the photon via kinetic mixing $\epsilon F^{\mu\nu} F'_{\mu\nu}$, allowing photon–dark photon oscillation. This motivates the light shining through the wall (LSW) studies aiming to detect the dark photon based on the oscillation of the photon into a dark photon, which is transmitted through the wall and subsequently converted back into the photon to be detected at the photodetector (PD) located behind the wall [13–16]. On the other hand, the ALP couples with the photon via the interaction term $g_{a\gamma\gamma} a F^{\mu\nu} \tilde{F}_{\mu\nu}$ or $g_{a\gamma\gamma} a \vec{E}.\vec{B}$, which enables the axion conversion into the photon via strong magnetic field [12].

Although the existence of dark photons and axions is often considered separately, their interaction with the photon implies that both of them are connected to each other [17,18]. In fact, this has been proposed to explain several physical observations. For instance, a model containing both an ALP and dark photon can explain the compatibility between the observation of the vacuum polarization experiment PVLAS [19] and astrophysical constraints [20]. In addition, the 3.55 keV line in the spectra of Galaxy clusters was explained using the model with both the ALP and dark photon [21]. Consequently, the direct interaction involving the SM particles, dark photon, and ALP, $G_{a\gamma\gamma'} a F^{\mu\nu} \tilde{F}'_{\mu\nu}$, is allowed in this kind of model. The presence of this coupling, in addition to the usual kinetic mixing between photon and dark photon $\epsilon$, opens up a new strategy to probe the dark photon.

Recently, Ref. [22] proposed a novel method to detect the dark photon based on its decay product. Instead of converting back into the photon, the transmitted dark photon decays into an axion and a photon. The proposal relies on the detection of the secondary photon in the final state. The main difference between this scenario and typical LSW experiments is that the frequency of the signal photon is lower than the original photon source.

As we will see below, this only requires a slight modification to the existing LSW experiments. On the other hand, the LSW experiments to probe the dark photon from photon–dark photon oscillation can also be implemented in gravitational wave (GW) experiments [23]. Therefore, in the presence of the new coupling $G_{a\gamma\gamma'}$, one can modify the corresponding setup to accommodate this coupling, which is the main objective of this paper.

---

*setyo13nugros@ntnu.edu.tw

Thus, our proposal provides a complementary method of dark photon search even though the smoking gun of the dark photon signal is different from that of photon–dark photon oscillation. Actually, the idea of looking for dark sector particles in laser interferometry experiments has been studied in [24–57].

The rest of this paper is organized as follows. In Sec. II, we give a brief review of the important aspects of dark photon detection using the secondary photon proposed in [22]. In Sec. III, we examine the implementation of secondary photon search at GW detectors. We discuss the estimated sensitivity in Sec. IV. Our summary and conclusions are presented in Sec. V.

## II. THE MODEL

We briefly review the model and the proposed experimental setup given in [22]. It contains a dark photon $\gamma'$ and a stable ALP $a$ residing in the dark sector. The dark photon kinetically mixes with the photon $\gamma$, while the ALP couples to both the dark photon and the photon. Because of the mixing, any light sources can be converted into a dark photon, which subsequently decays into an ALP and photon when $m_{\gamma'} > m_a$, see Fig. 1.

The proposed experimental setup relies on this decay aiming at the detection of the secondary photon, whose frequency is smaller than the photon from the light source. This secondary photon is called the subfrequency photon. In effective Lagrangian language, the relevant interaction is given by

$$\mathcal{L} \sim -\frac{\epsilon}{2} F_{\mu\nu} \tilde{F}'^{\mu\nu} + \frac{G_{a\gamma\gamma'}}{2} a F_{\mu\nu} \tilde{F}'^{\mu\nu}, \quad (2.1)$$

where $F_{\mu\nu} = \partial_\mu A_\nu - \partial_\nu A_\mu$ and $\tilde{F}'^{\mu\nu}$ stand for the field strength of the photon field $A_\mu$ and the dark photon field $A'_\mu$, respectively. To determine the number of the subfrequency signal photon, one needs the dark photon decay rate, which is given by [17]

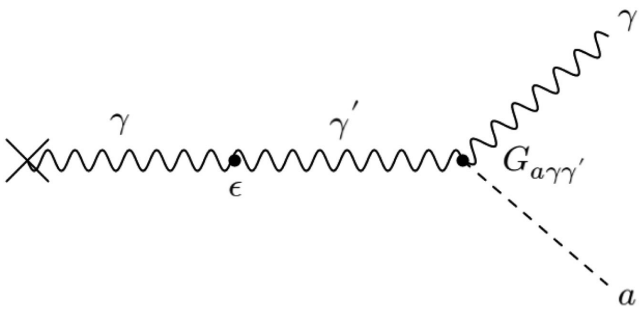

FIG. 1. The production of dark photon $\gamma'$ via kinetic mixing with the photon followed by its decay into a much lighter ALP $a$ and subfrequency photon $\gamma$.

$$\Gamma_{\gamma' \to a\gamma} = \frac{G^2_{a\gamma\gamma'}}{96\pi} m^3_{\gamma'} \left(1 - \frac{m^2_a}{m^2_{\gamma'}}\right)^3. \quad (2.2)$$

For a given light source, with power $P$ and frequency $\omega$, the flux of the subfrequency signal photon $N_{\text{sub}}$ having traversed a distance L is

$$N_{\text{sub}} = \epsilon^2 \left(1 - \exp\left[-\frac{m_{\gamma'} \Gamma_{\gamma' \to a\gamma} L}{\sqrt{\omega^2 - m^2_{\gamma'}}}\right]\right) N_\gamma, \quad (2.3)$$

where $N_\gamma = P/\omega$ is the photon number emitted from the source in units of hertz. In addition, Ref. [22] imposed the following conditions:

$$m_a \ll m_{\gamma'} < \omega,$$
$$1 \gg \frac{m_{\gamma'} \Gamma_{\gamma' \to a\gamma} L}{\sqrt{\omega^2 - m^2_{\gamma'}}}, \quad (2.4)$$

which states that the dark photon originated from the laser and it decays into a much lighter ALP and subfrequency photon. This implies that the subfrequency photon would have a sufficient number of events and could be detected using a typical photodetector. If the ALP mass is comparable to the dark photon mass $m_a \sim m_{\gamma'}$, the subfrequency photon would be too feeble to be detected at the optical photodetector considered here [22]. The last condition is to ensure that the signal photon increases linearly with L, which translates into the following bound:

$$L \ll \frac{\sqrt{\omega^2 - m^2_{\gamma'}}}{m^4_{\gamma'}} \left(\frac{G^2_{a\gamma\gamma'}}{96\pi}\right)^{-1}. \quad (2.5)$$

As a result, the relation between the number of signal photon flux $N_{\text{sub}}$ and the number of initial photon flux $N_\gamma$ can be written as

$$\frac{N_{\text{sub}}}{N_\gamma} = \frac{K^2}{96\pi} \frac{m^4_{\gamma'}}{\sqrt{\omega^2 - m^2_{\gamma'}}} L, \quad (2.6)$$

where $K$ is the product of two portal couplings,

$$K \equiv \epsilon G_{a\gamma\gamma'}. \quad (2.7)$$

As long as the bound on Eq. (2.5) is satisfied, the number of the signal photon scales linearly with L.

To detect the subfrequency photon, Ref. [22] suggests the experimental setup shown in Fig. 2. An optical laser is employed as a light source, which enters the Fabry-Prot (FP) cavity to amplify the laser beam, allowing the enhancement of dark photon transition rate with the amplification factor

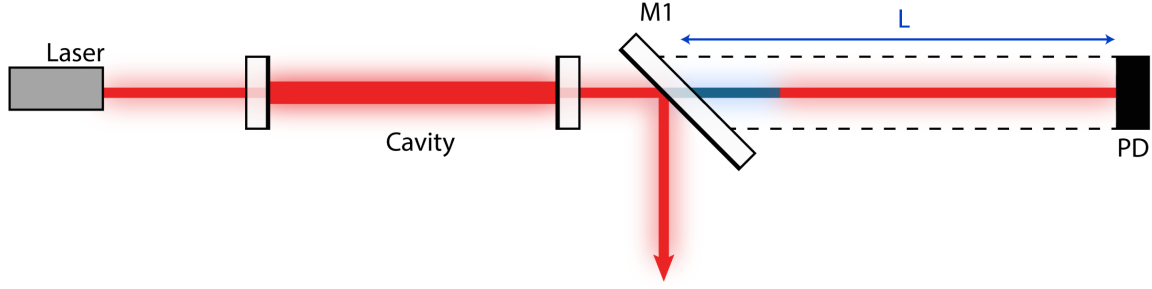

FIG. 2. The proposed experiment in an optical cavity in [22]. The photon (red line) propagates to the cavity from the laser source. Subsequently, it is reflected by the mirror M1 where only the dark photon (blue line) penetrates the mirror and decays into a subfrequency photon. The waveguide (dashed line) reflects the scattered subfrequency photon to be detected at the PD.

$$\eta_{\text{cav}} = \frac{N_{\text{pass}} + 1}{2}, \quad (2.8)$$

where $N_{\text{pass}}$ stands for the number of the beam reflection inside the cavity. A mirror M1 placed behind the cavity reflects all the outgoing laser beam from the cavity, while the dark photon is transmitted through the mirror. The signal photon resulting from the decay of the dark photon is collected by a photodetector located behind the mirror. Moreover, a waveguide (WG) is installed between the mirror and the photodetector to ensure all the signal photons would reach the detector.

Furthermore, the detection of the subfrequency signal photon requires a high efficiency detector as well as a low dark counting rate. Photomultiplier tubes and transition-edge sensors (TESs) are known devices sensitive to low photon counting. In a typical photodetector, the signal-to-noise ratio (SNR) of the single photon detection is given by

$$\text{SNR} = \frac{N_s \sqrt{t_s}}{\sqrt{N_s + N_d}}, \quad (2.9)$$

where $N_s$, $N_d$, and $t_s$ denote the number of signal photon, the number of dark current noise, and the total measurement time, respectively. For a given detector with the average efficiency $\bar{\eta}_{\text{eff}}$, the number of the detectable signal photon is

$$N_s = \eta_{\text{kin}} \eta_{\text{cav}} \bar{\eta}_{\text{eff}} N_{\text{sub}} = \eta_{\text{kin}} \eta_{\text{cav}} \bar{\eta}_{\text{eff}} \frac{K^2}{96\pi} \frac{m_{\gamma'}^4}{\sqrt{\omega^2 - m_{\gamma'}^2}} L N_\gamma. \quad (2.10)$$

Here, $\eta_{\text{kin}}$ stands for the probability of the signal photon arriving at the detector, which is optimized by using a waveguide. Several remarks are in order regarding Eq. (2.10). First, there is no perfect waveguide that collects all the photons. This would cause the loss of the signal photon inside the waveguide. Therefore, the probability of the subfrequency photon reaching the detector $\eta_{\text{kin}}$ becomes [22]

$$\eta_{\text{kin}} = \frac{1}{L} \frac{1}{\sqrt{\omega^2 - m_{\gamma'}^2}} \int_0^L d\ell \int_{E(\theta_{\text{lab}}=\pi/2)}^{E(\theta_{\text{lab}}=0)} dE \, \mathcal{R}(\theta_{\text{lab}}(E), \ell). \quad (2.11)$$

Here, $\mathcal{R}(\theta_{\text{lab}}(E), \ell)$ denotes the surviving fraction of the photon after passing the waveguide given by [22]

$$\mathcal{R}(\theta_{\text{lab}}(E), \ell) = r(E)^{\frac{1}{2} + \frac{\ell}{2R} \tan \theta_{\text{lab}}}, \quad (2.12)$$

where R and $r$ are the waveguide's radius and the reflectivity of the mirror in the waveguide, respectively. We set R = 8.75 mm equal to the radius of the lens utilized in the ALPS II experiment [14]. The polar angle of the signal photon in the lab frame $\theta_{\text{lab}}$ depends on its energy $E$ [22],

$$\theta_{\text{lab}}(E) = \cos^{-1} \frac{2E\omega - m_{\gamma'}^2}{2E\sqrt{\omega^2 - m_{\gamma'}^2}}. \quad (2.13)$$

Second, the detector efficiency $\eta_{\text{eff}}$ depends on the energy of the incident signal photon [22,58]. Thus, the averaged detector efficiency defined as the ratio between $\eta_{\text{eff}}$ and the probability of the signal photon arriving at the photodetector $\eta_{\text{kin}}$ is given by

$$\bar{\eta}_{\text{eff}} = \frac{1}{\eta_{\text{kin}}} \frac{1}{L} \frac{1}{\sqrt{\omega^2 - m_{\gamma'}^2}}$$
$$\times \int_0^L d\ell \int_{E(\theta_{\text{lab}}=\pi/2)}^{E(\theta_{\text{lab}}=0)} dE \, \eta_{\text{eff}}(E) \mathcal{R}(\theta_{\text{lab}}(E), \ell). \quad (2.14)$$

To demonstrate these two effects, we reproduce the sensitivity for ALPS II within the search of the subfrequency photon in Fig. 3 [22]. In addition, we also plot the sensitivity of the waveguide with L = 10 m. In the case of a perfect waveguide, the sensitivity scales linearly with $\sqrt{L}$, as can be seen from the three different dashed lines associated with different L and Eq. (2.9). In contrast, for a waveguide with $r = 98.5\%$, the sensitivity only improves linearly with $\sqrt{L}$ in the low mass region below 0.01 eV, where the sensitivity is fairly poor. In the higher mass regime, where the sensitivity achieves its maximum value, we see the dependence on L is very weak. The photon loss increases as the waveguide becomes longer. From here on, we take $L = 10$ m in our subsequent calculation to get the sensitivity in the GW experiments.

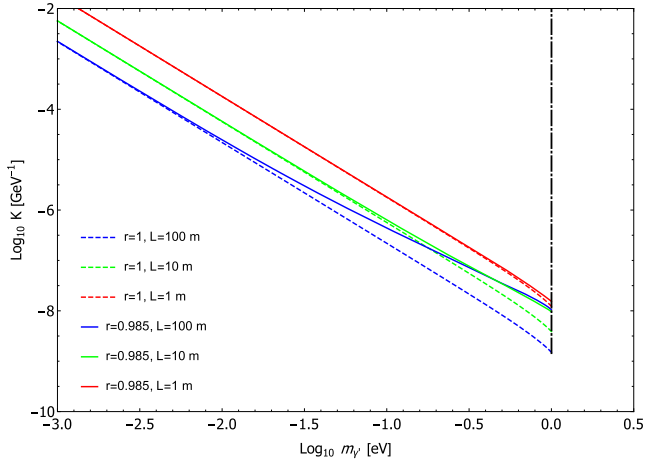

FIG. 3. The sensitivity of the ALPS II experiments with different waveguide configurations. The dashed lines indicate $r = 1$ (perfect waveguide), while the solid lines describe $r = 98.5\%$. We set the cutoff at $m_{\gamma'} = 0.99\omega$, which is indicated by the vertical dot-dashed line. We take 20 days of observation time in all cases.

## III. THE PROPOSED SETUP IN GW DETECTORS

The detection of the subfrequency photon can also be implemented in GW experiments. A typical GW experiment employs the Michelson interferometer to detect the spacetime fluctuation by measuring the differential arm length $\Delta\ell$ induced by the GW. The simplified version of the Michelson interferometer is depicted in Fig. 4.

In GW experiments, a laser beam is used as the light source. The beam splitter divides the laser beam into two perpendicular optical paths in the x and y direction with the same intensity. In the x direction, the beam enters the FP cavity formed by intermediate test mass (ITMX) and end test mass (ETMX). Inside the cavity, the laser beam receives power amplification, allowing them to produce a large number of photons. For a cavity with finesse $\mathcal{F}$, the number of reflections inside the cavity is

$$N_{\text{pass}} = \frac{2\mathcal{F}}{\pi}, \quad (3.1)$$

which can also be expressed as $N_{\text{pass}} = P_{\text{arm}}/P_{\text{in}}$. Here, $P_{\text{arm}}$ corresponds to the laser power inside the cavity, while $P_{\text{in}}$ is the laser power before entering the FP cavity. The same process also occurs in the y direction. Subsequently, the amplified beams coming from these cavities interfere with each other at the beam splitter (BS), producing the interference fringe to be detected at the PD, see Fig. 4. The observed interference pattern depends on the difference between these optical paths. Consequently, the measured change of the interference pattern is related to the differential arm length $\Delta\ell$ given by

$$\Delta\ell = \Delta\ell_x - \Delta\ell_y. \quad (3.2)$$

Since the GW experiments utilize the FP cavity, one expects the photon–dark photon conversion to take place inside the cavity. Comparing the setup in Fig. 2 with the GW experimental setup in Fig. 4, one sees that the end test mass (ETM) plays a similar role as M1. Having penetrated the ETM, the dark photon would decay into axion and signal photon after propagation of distance L. To detect the signal photon, we propose to install another photodetector (TES) equipped with a waveguide behind the ETM. In this paper, we will use a tungsten TES since it was reported to have the best efficiency $\eta_{\text{eff}}$ in the relevant wavelength of the subfrequency photon [22].

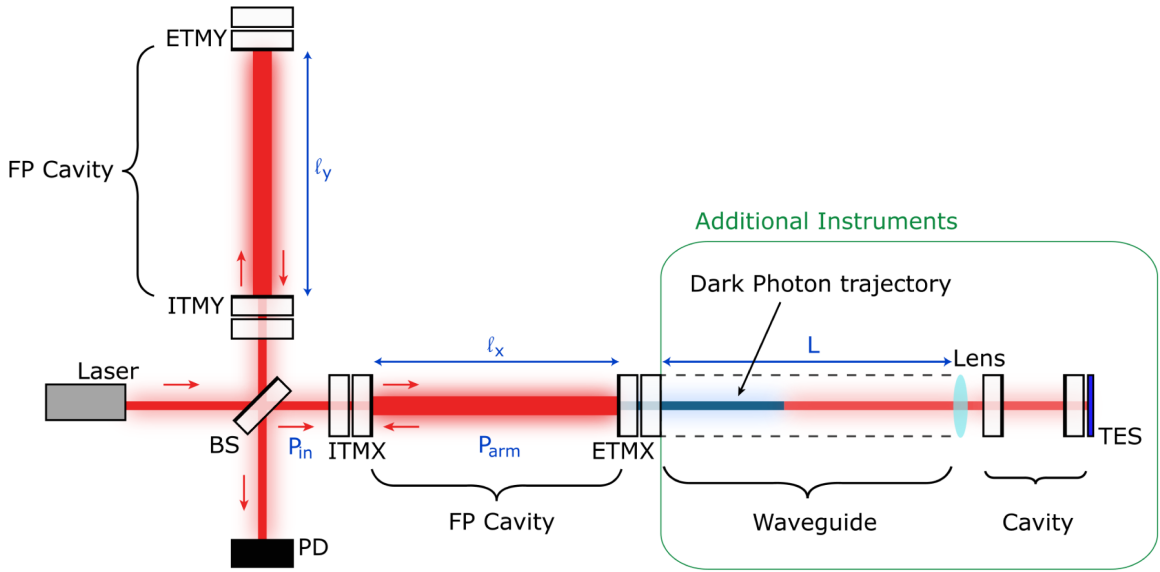

FIG. 4. The schematic of the experiment in the interferometric gravitational wave detector. We propose additional instruments aligned with the main beam in the FP cavity. The main feature is similar to [22], with an additional cavity to increase the number of subfrequency photons. Here, the length of the second cavity is negligible compared to the waveguide.

TABLE I. Parameters of the experiments used in our calculation. The references for the primary FP cavity and laser setup of each experiment are presented in the table. In addition, we adopt the ALPS II cavity finesse to our proposed second cavity for GW experiments.

| Parameters | ALPS II [14] | aLIGO [62] | KAGRA [63–65] | ET-HF [59–61] | ET-LF [59–61] |
|---|---|---|---|---|---|
| $\mathcal{F}_{\text{Cav}}$ | 7853 | 450 | 1550 | 880 | 880 |
| $P_{\text{in}}$ (W) | 30 | 2600 | 412 | 5355 | 32 |
| $\omega$ (eV) | 1.165 | 1.165 | 1.165 | 1.165 | 0.799 |
| $L$ (m) | 100 | 10 | 10 | 10 | 10 |
| $N_d$ (Hz) | $10^{-6}$ | $10^{-6}$ | $10^{-6}$ | $10^{-6}$ | $10^{-6}$ |
| $\mathcal{F}_{\text{Cav}}^{\text{WG}}$ | ⋯ | 7853 | 7853 | 7853 | 7853 |

Moreover, we suggest to include another cavity in this additional setup to increase the subfrequency photon signal with an amplification factor

$$\eta_{\text{cav}}^{\text{WG}} = \frac{1}{2}\left(\frac{2\mathcal{F}_{\text{WG}}}{\pi} + 1\right), \quad (3.3)$$

where $\mathcal{F}_{\text{WG}}$ denotes the finesse of the second cavity [23]. It will be placed next to the waveguide along with a lens to focus the signal photon. This is done to prevent the loss of unparalleled subfrequency photon during the signal amplification if it is installed along with the waveguide.

This slight modification will not disturb the main experimental setup to detect the GWs. Conversely, this would allow GW experiments to perform a search of new physics along with the detection of GWs. Notice that, although there will be a vacuum chamber installed surrounding the ETM, we can still conduct the experiment since only the dark photon signal will leave the main cavity and enter the waveguide. In addition, all the additional instruments should be confined inside a vacuum chamber to minimize the noise.

Taking into account the inclusion of the second cavity, the total detectable number of the subfrequency signal photon becomes

$$N_s^{\text{tot}} = \eta_{\text{kin}}\eta_{\text{cav}}\bar{\eta}_{\text{eff}}\eta_{\text{cav}}^{\text{WG}}\left(\frac{K^2}{96\pi}\frac{m_{\gamma'}^4}{\sqrt{\omega^2 - m_{\gamma'}^2}}L\right)N_\gamma, \quad (3.4)$$

where in this setup, L is taken as the distance between the ETM and the end of the waveguide since we consider dark photon decay occurs only inside the waveguide, as shown in Fig. 4.

To achieve high sensitivity measurement, the conventional gravitational wave experiments utilize high laser power in their design. A large number of photons allows them to precisely measure the change of the interference pattern. However, having a bunch of photons hitting the mirror would make the position of the mirror become unstable. This is due to the transferred momentum from the photons to the mirror, which further limits the ability to precisely determine the differential arm length of the interferometer. This is known as the radiation pressure noise, which dominates the experimental sensitivity in the low frequency regime.

To overcome this problem, the proposed third generation GW experiment, the Einstein Telescope (ET), plans to combine two separated interferometers. One interferometer is designed to be sensitive at the low frequency region (ET-LF), while the other one is constructed to achieve high sensitivity at the high frequency regime (ET-HF). In ET-LF, a lower laser power is used to suppress the radiation pressure noise. On the other hand, ET-HF exploits higher laser power in its structure. The final sensitivity is reached by combining the sensitivity of these two interferometers. This setup is known as a xylophone configuration [59–61].

We collect the optical properties of the interferometer in the current as well as the proposed GW experiments in Table I. The significant laser power enhancement in their optical cavity plays a crucial role in improving the number of subfrequency signal photon.

## IV. RESULTS AND DISCUSSION

We present the projected $1\sigma$ sensitivities of our proposal in Fig. 5, where we have taken 20 days of observation time. We set the cutoff $m_{\gamma'} = 0.99\omega$ in these curves to respect the limit in Eq. (2.5). The purple band is the reproduced sensitivity plot proposed by [22] in the ALPS II experiment with L = 100 m. The additional instruments in their setup consist of a mirror, waveguide, and photodetector (TES) as shown in Fig. 2.

Apart from ET-LF, there are more than 2 orders of magnitude enhancements in the sensitivity of the existing,

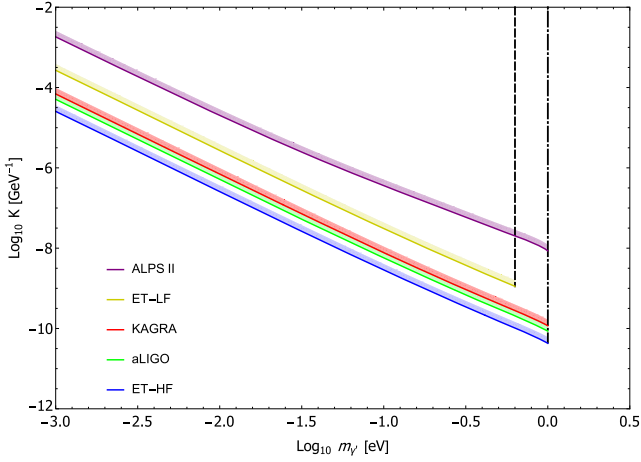

FIG. 5. The sensitivity of GW experiments on dark photon search using the subfrequency photon scenario. For all GW detectors, we use a waveguide with L = 10 m and r = 98.5%. In the case of ALPS II, L = 100 m. Both vertical black lines denote the cutoff at $m_{\gamma'} = 0.99\omega$.

as well as the proposed GW experiments compared to the ALPS II case. This is due to higher laser power, higher finesse in the cavity, and the inclusion of an additional WG cavity in these experiments. On the other hand, the ET-LF is only 1 order magnitude better than ALPS II due to the low laser power in their input. Both KAGRA and aLIGO utilize the high laser power, as well as high finesse in their FP cavities. However, the radiation pressure noise limits their detector sensitivity at the low frequency regime. In contrast, the ET-HF has the best sensitivity, since they are able to relieve this noise.

Since the detection method relies on the decay of the dark photon, the sensitivity is optimized when $m_{\gamma'}$ close to the laser frequency $\omega$, cf. Eq. (2.2). Furthermore, the experiments can reach higher dark photon mass by using a higher frequency laser source. One can improve the sensitivity by using higher laser power, as well as a higher finesse cavity, which are limited by the damage threshold of the mirror. In addition, suppressing the background noise in the detector, reducing the loss of the waveguide, and using the additional cavity with higher finesse in the detection regime would also enhance the sensitivity.

## V. SUMMARY AND CONCLUSIONS

A novel idea to detect a dark photon in LSW experiments has been proposed by Ref. [22]. This relies on the notion that any light sources can produce dark photon particles, which undergo decay into the subfrequency photon and ALP. We propose to extend this idea in the current and proposed GW experiments. We suggest to install additional devices, such as a photodetector, waveguide, and additional optical cavity in GW detectors, which allows us to detect this secondary photon with a smaller frequency. The capability of this simple setup to produce a new particle and its corresponding decay products, as well as detecting a new physics signal mimics high energy collider experiments such as the LHC. The smoking gun of the new physics signal in this mini-LHC is the detection of the secondary photon, whose frequency is smaller than the original laser input.

Our proposal is 2 orders of magnitude more sensitive than the original idea to be implemented at LSW experiments such as ALPS II [22]. We place new limits on the combined portal $K = \epsilon G_{a\gamma\gamma'}$, which are more stringent than the ones given in [22]. As pointed out in Ref. [22], these bounds cannot be drawn as a product of $\epsilon$ and $G_{a\gamma\gamma'}$ acquired independently, since it is quite model dependent.

In a model where both couplings are available, such as Ref. [18], the induced bound on $K$ is less than $10^{-7}$ for $m_{\gamma'} < 10^{-4}$ eV. This is obtained from the ALPS I experiment, which is sensitive to light dark photon mass below $10^{-3}$ eV. ALPS II would improve this limit by 3 orders of magnitude in the same mass range. From this point of view, our proposal can act as a complementary search of dark photons in the higher mass regime ($m_{\gamma'} > 10^{-3}$ eV) compared with ALPS I and ALPS II. As a closing remark, we do not take the induced limits from the sun and horizontal branch stars, since they suffer from astrophysical uncertainties as well being as model dependent [20,66–70]. In contrast, both ALPS I and ALPS II, as well as the GW experimental setup, are purely laboratory experiments operating in a well-controlled environment.

## ACKNOWLEDGMENTS

We would like to thank Dr. Jiheon Lee for insightful discussions. M. A. I. and H. T. W. are supported under Contracts No. MOST 106-2923 M-001-006-MY5 from the National Science and Technology Council, Taiwan and No. AS-TP-112-M01 from Academia Sinica. C. S. N. is supported by the Ministry of Science and Technology (MOST) of Taiwan under Grant No. MOST 111-2811-M-003-025.